\documentclass[%
 reprint,
superscriptaddress,
nofootinbib,
 amsmath,amssymb,
 aps,
]{revtex4-2}
\usepackage[english]{babel}
\usepackage{mathtools}
\usepackage{aps_macros}
\usepackage{pifont}
\usepackage{graphicx}
\usepackage{dcolumn}
\usepackage{bm}
\usepackage{hyperref}

\usepackage{subfigure}%
\usepackage{float}%
\hypersetup{
     colorlinks   = true,
     citecolor    = blue
}

\hypersetup{colorlinks=true, citecolor=blue, linkcolor = magenta}

\usepackage{orcidlink}

\begin{document}

\preprint{APS/123-QED}

\title{Nonradial oscillations of realistic anisotropic neutron stars: Polar modes}

\author{L.~M.~Becerra \orcidlink{0000-0002-3262-5545}}
\email{laura.becerra@umayor.cl}
\affiliation{Centro Multidisciplinario de F\'isica, Vicerrector\'ia de Investigaci\'on, Universidad Mayor,  Santiago de Chile 8580745, Chile}

\author{Jos\'e~F.~Rodr\'iguez-Ruiz \orcidlink{0000-0003-3627-5084}} 
\email{jrodriguez154@uan.edu.co}
\affiliation{Departamento de F\'isica, Universidad Antonio Nari\~no, Cra 3 Este \# 47A - 15, Bogot\'a D.C. 110231, Colombia}
 
\author{E.~A.~Becerra-Vergara \orcidlink{0000-0003-4848-1483}}
\email{eduar.becerra@correo.uis.edu.co}
\affiliation{Grupo de Investigaci\'on en Relatividad y Gravitaci\'on, Escuela de F\'isica, Universidad Industrial de Santander A. A. 678, Bucaramanga 680002, Colombia}

\author{F.~D.~Lora-Clavijo \orcidlink{0000-0003-4613-2917}} 
\email{fadulora@uis.edu.co}
\affiliation{Grupo de Investigaci\'on en Relatividad y Gravitaci\'on, Escuela de F\'isica, Universidad Industrial de Santander A. A. 678, Bucaramanga 680002, Colombia}

\date{\today}

\begin{abstract}

In this work, we study the polar perturbations of static, spherically symmetric neutron stars with anisotropic pressure in full general relativity, including linear-order perturbations of both the metric and the fluid. We calculate the $f$-mode frequencies and the corresponding damping times using a consistent treatment of the perturbation of the radial vector $k^\alpha$. In particular, its Lagrangian perturbation $\Delta k^\alpha$ is determined by the metric perturbations and the fluid Lagrangian displacement and is constrained to the $(\tilde{u},\tilde{k})$ plane, where $\tilde{u}^\alpha$ is the normalized fluid four-velocity. This constraint introduces an additional dynamical degree of freedom into the perturbation equations. Considering three equations of state and the Horvat and Bowers-Liang prescriptions for pressure anisotropy, we find that the $f$-mode frequency increases with stellar mass, ranging from $1$ to $3$~kHz, while the damping time decreases, ranging from $0.5$ to $1.25$~s. Increasing anisotropy, in the sense of tangential pressure exceeding radial pressure, generally lowers the oscillation frequency, while its effect on the damping time depends on the anisotropy prescription: the damping time decreases with increasing anisotropy for the Horvat model but increases with increasing anisotropy for the Bowers-Liang model. We further find quasi-universal relations between the real and imaginary parts of the $f$-mode frequency, $M\omega_R$ and $M\omega_I$, and the stellar compactness $\mathcal{C}=M/R$, which are largely insensitive to the equation of state. Polynomial fits to these relations achieve an accuracy better than $10\%$, providing a simple phenomenological framework for constraining neutron-star pressure anisotropy through future asteroseismology observations.

\end{abstract}

\maketitle


\section{\label{sec:intro} Introduction}
The detection of gravitational waves from binary neutron star mergers has ushered in a new era for the study of dense matter, providing new opportunities to constrain the equation of state (EOS) through complementary observables, including tidal deformabilities, stellar oscillations, and the post-merger gravitational-wave signal
\citep{2017PhRvL.119p1101A,2019JPhG...46k3002B,1998MNRAS.299.1059A}. In this context, relativistic asteroseismology provides a powerful tool for probing the internal structure of neutron stars through their quasinormal modes, whose frequencies and damping times depend on the stellar mass, radius, compactness, and EOS \citep{1967ApJ...149..591T,1985ApJ...292...12D,1983ApJS...53...73L,1999LRR.....2....2K,1991RSPSA.432..247C}.

Most of these studies  assume that the stellar fluid is isotropic. However, various physical processes, such as ultra-intense magnetic fields \citep{2012MNRAS.427.3406F,2015MNRAS.447.3278B}, superfluidity \citep{sokolov1980phase,1998NuPhB.531..478C}, exotic phases of matter \citep{2019PhRvD.100j3006B,PhysRevLett.29.382} or nuclear interactions at high densities \citep{1975ARA&A..13..335C,2000PhR...328..237H}, can lead to differences between the radial and tangential pressures, resulting in anisotropic stellar configurations \citep{1974ApJ...188..657B,1997PhR...286...53H}. Recent studies have shown that pressure anisotropy can significantly modify the macroscopic properties of neutron stars, including the mass-radius relation, tidal deformability, moment of inertia, and compactness \citep{2025PhRvD.111j3005B,2024PhRvD.109d3025B,2026PhRvD.113b3009B,2022MPLA...3750188P,2016CQGra..33i5005Y}.   

The study of nonradial oscillations in anisotropic stars has experienced significant progress in recent years, progressing from initial analyses using the Cowling approximation \citep{PhysRevD.85.124023} to fully relativistic treatments that consistently account for the coupling between matter and spacetime \citep{2026PhRvD.114b4028R}. 
The nonradial oscillation spectrum comprises the $g$-, $f$-, $p$-, and $w$-mode families, whose properties encode valuable information about the internal structure and macroscopic characteristics of neutron stars. In general relativity, these oscillations are described as quasinormal modes with complex frequencies, with the imaginary part characterizing their damping due to gravitational-wave emission.
In particular, the $w$ modes, identified by \citet{1992MNRAS.255..119K}, correspond primarily to spacetime oscillations and are characterized by relatively high frequencies and short damping times. 
Subsequent developments within the framework of full General Relativity for polar modes have shown that anisotropy can significantly modify the frequencies of fundamental and pressure modes, and may even trigger pressure-mode instabilities for sufficiently large anisotropies, a phenomenon not observed in isotropic stars \citep{Mondal:2023wwo,2024PhRvD.110h3020L}. More recently, the extension of the formalism to the study of polar $w$-modes has shown that anisotropy also modifies these predominantly spacetime oscillations and their associated quasi-universal relations \citep{2026PhRvD.114d4052Y}. These results have shown the sensitivity of nonradial oscillations to pressure anisotropy and establish them as a promising tool for probing the structure of anisotropic stars and, ultimately, extracting information about the physics of ultradense matter from future gravitational-wave observations. A consistent treatment of anisotropy under perturbations is therefore essential for fully characterizing these oscillations.

The aforementioned studies employ well-established prescriptions for pressure anisotropy, such as the Bowers--Liang \citep{1974ApJ...188..657B} and Horvat \citep{2011CQGra..28b5009H} models. However, because these models specify anisotropy only at equilibrium, they do not prescribe how its direction responds to perturbations, leaving the perturbation problem underdetermined. In our previous work \citep{2026PhRvD.114b4028R}, we resolved this for axial perturbations by requiring the Lagrangian perturbation of the anisotropy vector $k^\alpha$ to lie in the $(\bar{u},\bar{k})$ plane, and being consistent with its normalization and orthogonality conditions. This prescription is motivated by the physical assumption that the anisotropy is tied to the matter and advected with the fluid. We showed that this treatment introduces an  intrinsic coupling between matter and spacetime perturbations in the axial sector, demonstrating that anisotropy can affect modes beyond those traditionally regarded as purely gravitational. 

For this paper, we extend this treatment to polar perturbations by considering the anisotropy models proposed by Bowers-Liang and Horvat, in combination with realistic state equations consistent with observations of massive pulsars such as PSR J0348+0432 \citep{2019ApJ...887L..21R,2019ApJ...887L..24M} and PSR J0740+6620  \citep{2021ApJ...918L..27R,2021ApJ...918L..28M}, as well as with the constraints imposed by the gravitational wave events GW170817 \cite{2017PhRvL.119p1101A} and GW190814 \cite{LIGOScientific:2020zkf}. Using a relativistic formulation based on an anisotropic energy-momentum tensor, we obtain the system of equations that describes the polar perturbations and allows us to determine the frequencies of their quasinormal modes. In particular, we analyze the $f$-modes, focusing on the influence of stellar mass and the degree of anisotropy on their oscillation frequencies and damping times.

This work is organized as follows. In Sec.~\ref{sec:Pert}, we briefly review the formalism for computing nonradial pulsations of spherically symmetric anisotropic neutron stars, with further details provided in \cite{2026PhRvD.114b4028R}. We then derive the equations governing the polar perturbations and describe the equations of state for the radial pressure and pressure-anisotropy prescriptions adopted in our analysis. In Sec.~\ref{sec:numerical}, we describe our numerical implementation, including the integration scheme, the treatment of the stellar surface, and the matching of the boundary conditions required to determine the complex quasinormal-mode frequencies. Section~\ref{sec:polarmodes} presents our main results for the $f$-modes of the polar perturbation, examining the dependence of their oscillation frequencies and degree of anisotropy. We also provide simple polynomial fits relating the mode frequencies to the stellar compactness and anisotropy parameter, and compare our results with previous formalisms, particularly those developed in \cite{2024PhRvD.110h3020L, 2026PhRvD.113d3025G}. Finally, in Sec.~\ref{sec:discussion}, we discuss our findings and summarize our main conclusions. Appendix~\ref{app:I} provides explicit expressions for the static background configuration and the polar perturbation equations, including their series expansions near the stellar center.

\section{\label{sec:Pert} Perturbations of static and spherically symmetric spacetime}
The line element of the static, spherically symmetric background spacetime in polar-areal coordinates is given by
\begin{equation}
    ds^2 = -A(r)dt^2 + B(r)^{-1}dr^2 + r^2(d\theta^2 + \sin^2\theta d\phi^2),
\end{equation}
which is sourced by an anisotropic fluid whose energy-momentum tensor is
\begin{equation}
    \bar{T}_{\alpha \beta} = (\bar{\epsilon} + \bar{P}_\perp) \bar{u}_\alpha \bar{u}_\beta
    + \bar{P}_\perp \bar{g}_{\alpha \beta} + (\bar{P} - \bar{P}_\perp) \bar{k}_\alpha \bar{k}_\beta,
\end{equation}
where $\bar{\epsilon}$, $\bar{P}$, and $\bar{P}_\perp$ denote energy density, radial, and tangential pressures, respectively. The normalized fluid four-velocity is $\bar{u}^\alpha = \delta^\alpha_t / \sqrt{A}$, and $\bar{k}^\alpha = \delta^\alpha_r / \sqrt{B}$ is the unit spacelike radial vector orthogonal to the flow ($\bar{u}_\alpha \bar{k}^\alpha = 0$).

Linear perturbations $h_{\alpha\beta} = g_{\alpha\beta} - \bar{g}_{\alpha\beta}$ are decomposed into spherical harmonics $Y^{\ell m}(\theta, \phi)$ and decouple into independent axial (odd-parity) and polar (even-parity) sectors \citep{Regge:1957td}. Without loss of generality, we work in the Regge-Wheeler gauge and restrict to axisymmetric ($m=0$) modes, following the general relativistic formulation developed in \citet{2026PhRvD.114b4028R}.

The fluid displacements are described by the Lagrangian displacement field $\xi^\alpha$, linking Eulerian ($\delta$) and Lagrangian ($\Delta$) variations via $\Delta = \delta + \mathcal{L}_\xi$. Preserving the normalization and orthogonality conditions under perturbation, namely $\Delta (g_{\alpha\beta }u^\alpha u^\beta) = \Delta(g_{\alpha\beta }k^\alpha k^\beta) = \Delta (g_{\alpha\beta }u^\alpha k^\beta) = 0$, yields the covariant relations for the Lagrangian perturbations of $u^\alpha$ and $k^\alpha$:
\begin{align}
    \Delta u^\alpha &= \frac{1}{2}\bar{u}^\alpha \bar{u}^\mu \bar{u}^\nu \Delta g_{\mu\nu}, \label{eq:Deltau} \\
    \Delta k^\alpha &= \bar{u}^\mu \bar{k}^\nu \Delta g_{\mu\nu} \bar{u}^\alpha - \frac{1}{2} \bar{k}^\mu \bar{k}^\nu \Delta g_{\mu\nu} \bar{k}^\alpha, \label{eq:Deltak}
\end{align}
where we have assumed that the Lagrangian perturbation of $k^\alpha$ lies in the $(\bar{u},\bar{k})$ plane \citep{2026PhRvD.114b4028R}.
The corresponding Eulerian perturbations follow directly as $\delta u^\alpha = \Delta u^\alpha - \mathcal{L}_\xi \bar{u}^\alpha$ and $\delta k^\alpha = \Delta k^\alpha - \mathcal{L}_\xi \bar{k}^\alpha$. 
In contrast, other works, e.g. \cite{PhysRevD.85.124023, Mondal:2023wwo, 2024PhRvD.110h3020L}, have assumed that the Eulerian perturbation of $ k^\alpha$ lies in the  $(\bar{u},\bar{k})$ plane. This assumption changes the dynamical character of the $\theta$-component of the displacement vector.

Combining these Eulerian variations with the thermodynamic perturbations $\delta\epsilon$, $\delta P$, and $\delta P_\perp$, the dynamics of the system are governed by the linearly perturbed Einstein field equations,
\begin{equation}
    \delta G_{\mu\nu} = 8\pi \delta T_{\mu\nu}.
\end{equation}
For a comprehensive derivation of these perturbation equations and the explicit  expressions for $\delta T_{\mu\nu}$, we refer the reader to \citet{2026PhRvD.114b4028R}.

\subsection{Polar Perturbations}
%
%
In the Regge-Wheeler gauge, the polar metric perturbations for axisymmetric modes ($m=0$) are expanded in terms of spherical harmonics $Y^{\ell 0}(\theta)$. 
Accordingly, the  metric perturbations takes the form:
\begin{equation}
(h^{\ell 0}_{\alpha\beta})^{\rm polar}=
\begin{bmatrix}
A\, H_0  & H_1 & 0 & 0 \\
H_1 & H_2/B & 0 & 0 \\
0 & 0 & r^2 K & 0 \\
0 & 0 & 0 & r^2 K \sin^2 \theta
\end{bmatrix}
Y^{\ell 0}(\theta).
\end{equation}
 Each perturbation function captures a distinct physical aspect of spacetime deformation: $H_0(r,t)$ represents the relativistic gravitational potential perturbation, $H_1(r,t)$ describes spacetime shear and frame-dragging driven by radial mass motion, $H_2(r,t)$ measures radial spatial curvature, and $K(r,t)$ tracks the fractional area deformation of concentric 2-spheres.

The spatial motion of the fluid is governed by the polar Lagrangian displacement vector,
\begin{equation}
    (\xi^\alpha)^{\rm polar} = \left[0,\, W, \, -V \,\partial_\theta, \, 0 \right] Y^{\ell 0}(\theta),
\end{equation} 
where $W(r,t)$ and $V(r,t)$ represent the radial and transverse displacement amplitudes, respectively. Applying the kinematic definitions in Eqs.~(\ref{eq:Deltau}) and (\ref{eq:Deltak}), the Eulerian perturbations of the fluid four-velocity $u^\alpha$ and the spacelike radial unit vector $k^\alpha$ take the form
\begin{equation}
    (\delta u^\alpha)^{\rm polar} = \frac{1}{\sqrt{A}}\left[ \frac{H_0}{2} , \,\partial_t W , - \partial_t V\,\partial_\theta , \, 0 \right] Y^{\ell 0}(\theta),
\end{equation}
and
\begin{equation}
    (\delta k^\alpha)^{\rm polar} = \sqrt{B}\left[ \frac{H_1}{A}+\frac{\partial_t W}{AB}, \, -\frac{H_2}{2}, \, - \partial_r V\,\partial_\theta , \, 0 \right] Y^{\ell 0}(\theta),
\end{equation}
respectively. Here, $\delta u^\alpha$ tracks the local fluid velocity field, whereas $\delta k^\alpha$ captures the geometric deformation of the surfaces of constant radius.

Consequently, the Eulerian perturbation of the energy-momentum tensor reduces to the symmetric matrix
\begin{equation}
(\delta T_{\alpha\beta}^{\ell 0})^{\rm polar}=
\begin{bmatrix}
s^{tt}& s^{Rt} & s^{Et} \partial_\theta & 0 \\
s^{Rt} &s^{L0} & s^{E1}\partial_\theta  & 0 \\
s^{Et} \partial_\theta   & s^{E1}\partial_\theta  & s^{T0} & 0 \\
0 & 0 & 0 & s^{T0}\sin^2\theta
\end{bmatrix} Y^{\ell 0}(\theta),
\end{equation}
whose individual components encode the perturbed energy density, anisotropic pressures, and momentum fluxes:
\begin{align}
    s^{tt} &= A(\delta \epsilon -\bar{\epsilon}H_0), \\
    s^{L0} &= \frac{1}{B}(\delta P +\bar{P}H_2), \\
    s^{T0} &= r^2(\delta P_\perp +\bar{P}_\perp K), \\
    s^{Rt} &= -\frac{1}{B}(\bar{\epsilon}+\bar{P})\partial_t W -\bar{\epsilon}H_1, \\
    s^{Et} &= r^2(\bar{\epsilon}+\bar{P}_\perp)\partial_t V, \\
    s^{E1} &= -r^2\bar{\sigma} \partial_r V.
\end{align}
Here, $\delta \epsilon$, $\delta P$, and $\delta P_\perp$ denote the Eulerian perturbations of the energy density, radial pressure, and tangential pressure, while $\bar{\sigma} \equiv \bar{P} - \bar{P}_\perp$ measures the background pressure anisotropy. 

Evaluating the difference between the transverse components of the Einstein field equations, $\delta G^\theta_\theta - \delta G^{\phi}_\phi = 8\pi (\delta T^\theta_\theta - \delta T^\phi_\phi)$, yields the algebraic constraint
\begin{equation}
    H_2 = H_0,
\end{equation}
showing that the radial curvature perturbation is fully determined by the metric potential $H_0$ in the absence of off-diagonal matter stress variations.

The remaining non-trivial components of the linearized Einstein field equations yield the following coupled metric-matter dynamic system:
\begin{widetext}
\begin{align}
    &B\partial_r H_1 - \partial_t H_0-\partial_t K + \left(\frac{BA'}{A}+B'\right) \frac{H_1}{2} = -16 \pi s^{Et}, \label{eq:EEa} \\
    &\partial_t\partial_r K + \left[\frac{BA'}{A}-B'-\frac{\ell(\ell+1)}{2r}\right]\frac{H_1}{r} -\frac{\partial_t H_0}{r} + \left(\frac{1}{r}-\frac{A'}{2A} \right)\partial_t K = -8\pi (s^{Rt} +H_1\bar{P} ), \label{eq:EEb} \\
    &\frac{\partial_t H_1}{A} + \partial_r K-\partial_r H_0 - \frac{A'}{A} H_0 = -16 \pi s^{E1}, \label{eq:EEc} \\
    &\partial^2_r K +\left(\frac{3}{r}+\frac{B'}{2B}\right) \partial_r K -\frac{\partial_r H_0}{r} - (\ell^2+\ell-2)\frac{K}{2r^2B} -\left(\frac{B'}{B}+ \frac{\ell(\ell+1)}{2rB} -\frac{1}{r}\right)\frac{H_0}{r} = \frac{8\pi \delta \epsilon}{B}, \label{eq:EEd} \\
    &\frac{\partial^2_t K}{AB} -\left(\frac{1}{r}+\frac{A'}{2A}\right) \partial_r K -\frac{\partial_t H_1}{rA} +(\ell^2+\ell-2)\frac{K}{2r^2B} +\frac{\partial_r H_0}{r}+\left(\frac{A'}{A} - \frac{\ell(\ell+1)}{2rB} +\frac{1}{r}\right)\frac{H_0}{r} = -\frac{8\pi \delta P}{B}. \label{eq:EEe}
\end{align}
\end{widetext}
Here, Eq.~(\ref{eq:EEa}) couples the metric shear $H_1$ to the transverse fluid momentum flux $s^{Et}$, while Eq.~(\ref{eq:EEb}) relates the time-varying spatial curvature gradient $\partial_t \partial_r K$ to the radial energy flux $s^{Rt}$. Equation~(\ref{eq:EEc}) serves as a constraint linking potential gradients to the anisotropic shear stress $s^{E1}$. Finally, the diagonal spatial components in Eqs.~(\ref{eq:EEd}) and (\ref{eq:EEe}) act as the relativistic Poisson equation for the energy density perturbation $\delta \epsilon$ and the wave equation for the radial pressure perturbation $\delta P$, respectively.

Furthermore, perturbing the energy-momentum conservation law governs the internal fluid dynamics:
\begin{widetext}
\begin{align}
    &\partial_t \delta \epsilon + (\bar{\epsilon}+\bar{P}_\perp) \left[\ell(\ell+1)\partial_t V +\partial_t K+\frac{2\partial_t W}{r}\right] + \frac{d\bar{\epsilon}}{dr}\partial_t W  
    + (\bar{P}+\bar{\epsilon})\left[\frac{\partial_t H_0}{2}+\partial_{tr} W-\left(\frac{A'}{2A} -\frac{4\pi r (\bar{\epsilon}+\bar{P})}{B}\right)\partial_t W\right] = 0, \label{eq:TOVa} \\
    &\partial_r \delta P + \frac{2\delta \sigma}{r} +\bar{\sigma}\partial_r K -\ell (\ell+1) \bar{\sigma}\partial_r V +(\delta P +\delta\epsilon)\frac{A'}{2A} + \left(\bar{P}+\bar{\epsilon}\right)\left(\frac{\partial_t H_1}{A}+\frac{\partial_{tt} W}{AB}-\frac{\partial_r H_0}{2}\right) = 0, \label{eq:TOVb} \\
    &\frac{\delta P_\perp}{r^2B}-\frac{(\bar{P}+\bar{\epsilon})H_0}{2r^2B} - \frac{(\bar{P}_\perp+\bar{\epsilon})\partial_{tt} V}{AB} -\partial_r \left(\bar{\sigma} \partial_r V \right)-\left(\frac{A'}{2A}+\frac{B'}{2B}+\frac{4}{r}\right)\bar{\sigma}\partial_r V = 0. \label{eq:TOVc}
\end{align}
\end{widetext}
Here, Equation~(\ref{eq:TOVa}) represents the perturbed energy continuity equation, and Equations~(\ref{eq:TOVb}) and (\ref{eq:TOVc}) correspond to the perturbed radial, and transverse Euler equations, respectively.

To close the system, we assume a barotropic EOS under adiabatic perturbations, where the Lagrangian variations of radial pressure ($\Delta P$) and energy density ($\Delta \epsilon$) are related via
\begin{equation}
    \Delta P = \frac{\Gamma_1 \bar{P}}{\bar{\epsilon} + \bar{P}} \Delta\epsilon,
\end{equation}
where the adiabatic index is defined as
\begin{equation}
    \Gamma_1 = \frac{\bar{\epsilon}+ \bar{P}}{\bar{P}}\left(\frac{\partial P}{\partial \epsilon}\right)_{\rm eq}.
\end{equation}
\subsection{Anisotropy models and equation of state}
We consider two standard prescriptions for pressure anisotropy, namely the Bowers-Liang \citep{1974ApJ...188..657B} and Horvat \citep{2011CQGra..28b5009H, PhysRevD.85.124023} models.
\begin{eqnarray}
\bar{\sigma}_{BL} &=& -\lambda_{BL} (\bar{\epsilon} + 3\bar{P})(\bar{\epsilon} + \bar{P}) \frac{r^2}{B} , \label{eq:sigma_BL}\\
\bar{\sigma}_H    &=&  -\lambda_H  \left(1-B\right) \bar{P} \, . \label{eq:Delta_p}
\end{eqnarray}
The corresponding dimensionless parameters, $\lambda_{BL}$ and $\lambda_H$, control the strength of anisotropy. Following \citep{2025PhRvD.111j3005B}, we restrict both parameters to $0\leq\lambda_{BL},\lambda_H<1$, ensuring physically well-behaved stellar configurations with monotonically decreasing radial pressure and finite radial and tangential  sound speed at the center.

We adopt the same three EOSs and numerical implementation adopted in \citet{2026PhRvD.114b4028R}: SLy4, GM1Y6, and QHC21, representing nucleonic, hyperonic, and hybrid matter, respectively. The EOSs are implemented using the Generalized Piecewise Polytropic (GPP) scheme of \citet{Boyle2020}, with a three-zone parametrization above nuclear saturation density, $\rho_s \simeq 2.4\times10^{14}\,\mathrm{g\,cm^{-3}}$, and a five-zone parametrization at lower densities. The corresponding parameters and numerical implementation are the same as those described in \citet{2024PhRvD.109d3025B} and \citet{2026PhRvD.114b4028R}.

\section{Numerical implementation}\label{sec:numerical}
The complete system of equations used in the numerical integration is presented in Appendix~\ref{app:I}. For completeness, we also include the equations describing the static, spherically symmetric anisotropic background, given by Eqs.~\eqref{eq:da}–\eqref{eq:dP}. Outside the stellar surface, where $P = P_\perp = \epsilon = 0$, these equations reduce to the Schwarzschild vacuum solution, $A = B = 1 - 2M/r$, where the stellar mass is given by
$M=[1-B(R)]R/2$. For the polar perturbations, we assume a harmonic time dependence of the form $e^{i\omega t}$ for all perturbed quantities, with the evolution governed by Eqs.~\eqref{eq:EEd}–\eqref{eq:TOVc}. The resulting system consists of six coupled first-order differential equations, Eqs.~\eqref{eq:polar_a}--\eqref{eq:polar_f}, together with the algebraic constraint given by Eq.~\eqref{eq:polar_g}.

The quasinormal modes correspond to the discrete set of complex frequencies, $\omega = \omega_R + i\,\omega_I$, that satisfy the appropriate physical boundary conditions: the perturbations must remain regular at the stellar center, the Lagrangian perturbation of the radial pressure, $\Delta P$, must vanish at the stellar surface, $r=R$, and the metric perturbation functions $H_0$, $H_1$, and $K$ must behave as purely outgoing waves in the asymptotic limit ($r \rightarrow \infty$).

All differential equations are integrated using a fourth-order Runge--Kutta scheme on a one-dimensional spherical grid extending from the stellar center ($r = 0$) to an outer boundary located at $r_{\rm max} = 50M$, with a uniform grid spacing of $\Delta x =10^{-4} M $.  The quasinormal-mode frequencies are then computed following the procedure introduced by \cite{1985ApJ...292...12D,2011ChPhB..20d0401L} for isotropic stars. The numerical procedure is summarized below. 

\begin{itemize}

\item \emph{Background integration.} The background stellar model is first obtained by integrating Eqs.~\eqref{eq:da}--\eqref{eq:dP} outward from the center,with the central energy density, $\bar{\epsilon}(0)=\bar{\epsilon}_c$, specified as the initial condition. The integration is terminated at the stellar surface, $r=R$, defined by the condition $\bar{\rho}(R)=10^4,{\rm g, cm^{-3}}$.

 \item \emph{Interior perturbations.} Inside the star, the perturbation equations, Eqs.~\eqref{eq:polar_a}--\eqref{eq:polar_f}, form a system of six coupled first-order differential equations for the vector of variables  $ \Psi=\{H_1,\,K,\,W,\,V,\,\tilde{X},\,\chi\}$,  while the metric perturbation $H_0$ is obtained algebraically from Eq.~\eqref{eq:polar_g}. The system is integrated outward from the stellar center to the matching radius, $r=R/2$. Regularity at the origin is imposed by means of a power-series expansion of the perturbation variables about $r=0$ (see Appendix~\ref{app:I}), which provides the initial conditions at $r=\Delta r$. Two linearly independent solutions are generated by choosing
\begin{equation}
V_c=1, \qquad
K_c=\pm\left(\bar{\epsilon}_c+\bar{P}_c\right),
\end{equation}
and are denoted by $\Psi_1$ and $\Psi_2$.
To impose the boundary conditions at the stellar surface, four additional linearly independent solutions, denoted by $\Psi_3$--$\Psi_6$, are obtained by integrating the same system inward from $r=R$ to the matching radius, $r=R/2$. Their boundary values at $r=R$ are prescribed as
\begin{equation}
\begin{aligned}
\{H_1,K,W,V,\tilde{X},\chi\} &=
\{1,0,0,0,0,0\},\\
&\quad \{0,1,0,0,0,0\},\\
&\quad \{0,0,1,0,0,0\},\\
&\quad \{0,0,0,1,0,0\},
\end{aligned}
\end{equation}
respectively. The matching conditions at $r=R/2$ require continuity of all perturbation variables, yielding
\begin{equation}
a_1\Psi_1+\Psi_2
=
a_3\Psi_3+a_4\Psi_4+a_5\Psi_5+a_6\Psi_6.
\end{equation}
The coefficients $a_i$ are determined by solving the resulting linear system. The perturbation vector at the stellar surface is then given by
$\Psi_s=a_1\Psi_1+\Psi_2.$

    \item \emph{Exterior perturbations.} In the vacuum exterior of the star, where $P=P_\perp=\epsilon=0$,  the fluid perturbations vanish identically, and the metric perturbations, $H_1$ and $K$, satisfy a closed system of two coupled first-order equations, which can be combined into the Zerilli equation:
    \begin{equation}
        \left(\frac{d^2}{dx^2}  + (V_z+\omega^2 )\right)Z= 0,
    \end{equation}
    where the tortoise coordinate is defined by $dx/dr = 1/\sqrt{AB}$ and the Zerilli potential is 
    \begin{align}
    & V_z =  \frac{18(1-2M/r)}{(3M+nr)^2} \times \nonumber \\
    & \left[ \frac{M^3}{r^3} + \frac{nM^2}{r^2} + \frac{n^2M}{3r} + \frac{n^2(1 + n)}{9} \right] \,,
\end{align}
with $2n=(\ell-1)(\ell+1)$. The Zerilli function, $Z$,  is related to the metric perturbations through
    \begin{align}
        Z &= \frac{(2 M - r) r}{3 M + n r}H_1 + \frac{r^2}{3 M + n r}K \\
        \frac{dZ}{dr} &= \frac{6 M^2 + 3 M n r + n (1 + n) r^2 }{(3 M + n r)^2}H_1  \\
        & + \frac{
 r (-3 M^2 - 3 M n r + n r^2) }{(2 M - r) (3 M + n r)^2}K \nonumber 
    \end{align}

The values of $H_1$ and $K$ obtained from $\Psi_s$ are used to evaluate $Z$ and $dZ/dr$ at $r=R$, providing the initial conditions for integration of the Zerilli equation to the outer grid boundary. In the asymptotic region, $r\rightarrow\infty$, the solution can be expressed as a superposition of incoming and outgoing waves:
 \begin{equation}
     Z \approx A_{\rm in} e^{i \omega x}\sum_j \alpha_j r^{-j} +  A_{\rm out } e^{-i \omega x}\sum_j \alpha_j r^{-j}  ,
 \end{equation}
where the first coefficients of the asymptotic expansion are
 \begin{align}
     \alpha_1 &= -\frac{i (n+1)}{\omega }\alpha_0\\
     \alpha_2 &= -\frac{n (n + 1)}{\omega^2} \alpha_0 + 
 \frac{i 3 M (n + 2)}{2n\omega} \alpha_0  .
 \end{align}

 The coefficients $A_{\rm in}$ and $A_{\rm out}$ are determined by matching the numerical solution to the asymptotic expansion at the outermost grid point. Quasinormal modes correspond to complex frequencies for which the incoming-wave component vanishes, $A_{\rm in}(\omega)=0$.

\end{itemize}

For real $\omega$ the perturbation equations are real, so that $A_{\rm out}(\omega)=A_{\rm in}^*(\omega)$; this property is exploited, following \cite{1985ApJ...292...12D,2011ChPhB..20d0401L}, to efficiently locate modes with $\omega_I \ll \omega_R$. The frequency dependence of $A_{\rm out}$ near a given mode is approximated by the quadratic polynomial
\begin{equation}
A_{\rm out}(\omega)=a_1+a_2\omega+a_3\omega^2,
\end{equation}
with complex coefficients $a_1$, $a_2$, $a_3$ determined by evaluating $A_{\rm out}$ at three real trial frequencies. The complex root of this polynomial corresponding to the quasinormal mode is then obtained via a Newton--Raphson algorithm. Because the accuracy of the quadratic approximation depends on the trial frequencies lying sufficiently close to the real part of the target mode, the frequency obtained from the initial interpolation is treated only as a first estimate: its real part is used to select a new set of three real frequencies, and the procedure is repeated until $\omega_R$ converges to the desired precision.

\begin{figure*}
    \centering
    \includegraphics[width=0.94\textwidth]{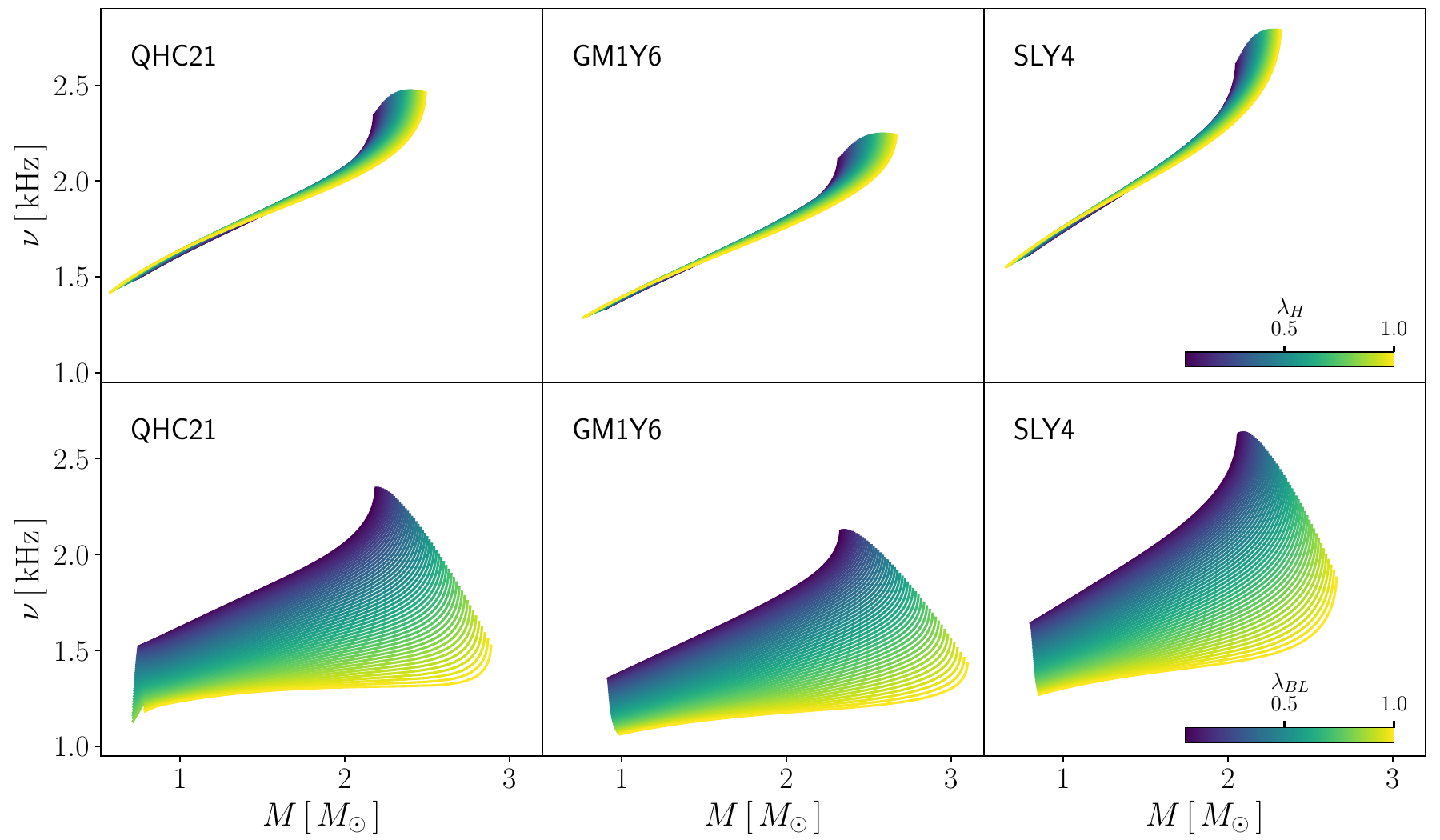}
    \caption{Oscillation frequency of the fundamental mode (f-mode) as a function of the stellar mass for different EOS and the Bowers-Liang (lower panel) and the Horvat (upper panel) anisotropy model. The color scale corresponds to the value of the anisotropic parameter.}
    \label{fig:omega}
\end{figure*}

\begin{figure*}
    \centering
    \includegraphics[width=0.94\textwidth]{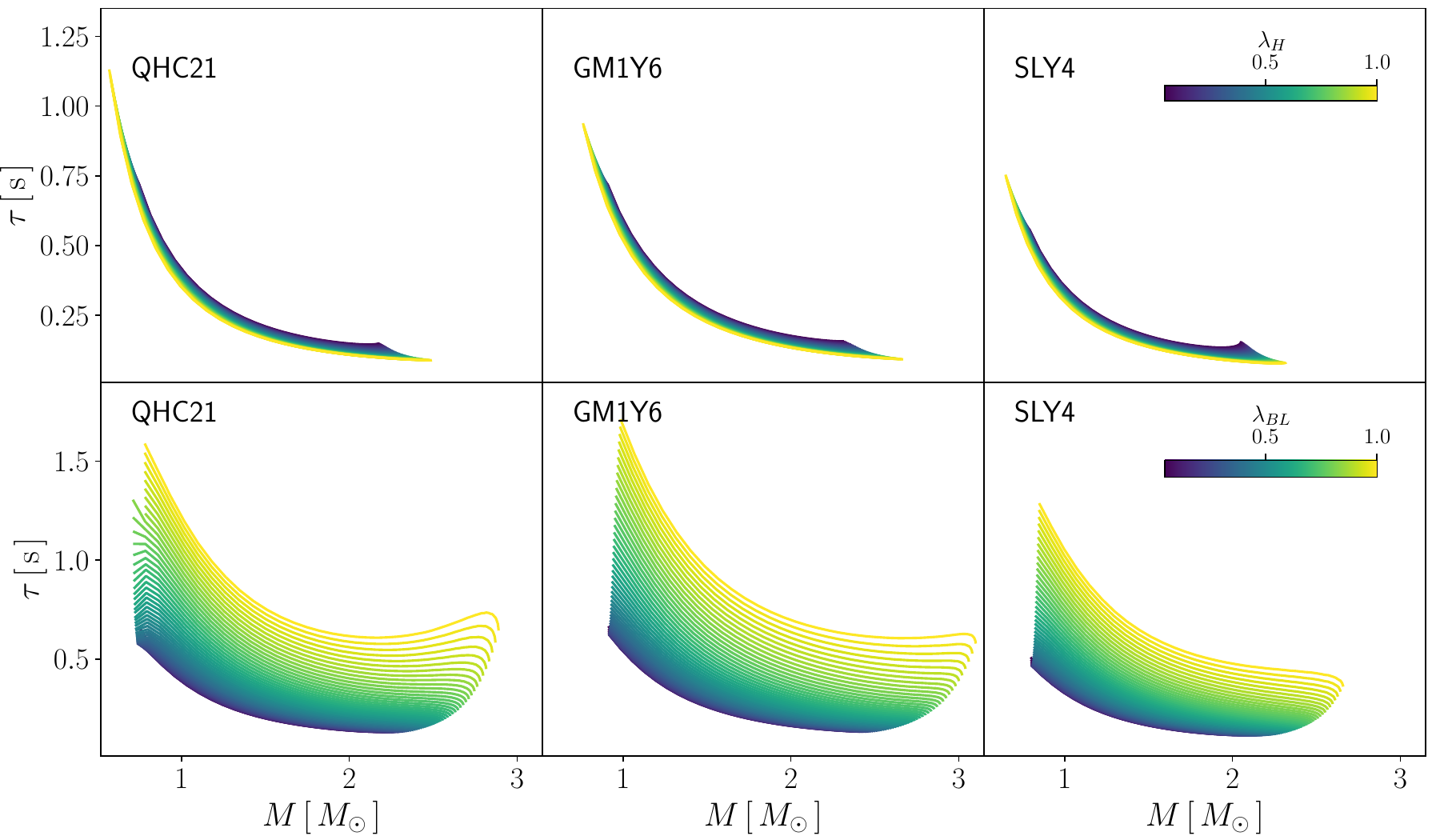}
    \caption{Damping time of the fundamental mode (f-mode) as a function of the stellar mass for different EOS and the Bowers-Liang (lower panel) and the Horvat (upper panel) anisotropy model. The color scale corresponds to the value of the anisotropic parameter.}
    \label{fig:tau}  
\end{figure*}

\begin{table*}\label{tab:fit_Momega}
    \begin{tabular}{c| ccccc}
     $\tilde{a}_{ij}$ & $ j=0 $ & $ j=1 $ & $ j=2$ & $ j=3 $ & \\ \hline\hline
         & \multicolumn{5}{c}{Hovart anisotropy model }  \\ 
    \hline
        $i=0$  & $0.00324\pm 0.00006$ & $-0.0035\pm 0.0005$ & $-0.0139 \pm 0.0011 $ &  $0.00893\pm 0.00071$ &\\ 
        $i=1$  & $0.2115\pm 0.0014$   & $0.0560 \pm 0.0113 $  & $0.2608 \pm 0.0245$ & $-0.1351 \pm 0.0153$ & \\ 
        $i=2$  & $1.802\pm 0.009$   & $-0.320 \pm 0.077$  & $-1.027 \pm 0.167$  & $0.317 \pm 0.105$ & \\ 
        $i=3$  & $-2.484\pm 0.018$   &  $1.975\pm 0.145$   &  $-0.307 \pm 0.314$  &  $0.485 \pm 0.197$ & \\ 
       \hline
        & \multicolumn{5}{c}{Bowers Liang anisotropy model }  \\ 
    \hline
        $i=0$  & $0.00324\pm 0.00006$  & $-0.023 \pm 0.002$   & $-0.022\pm0.003$    & $0.0047\pm 0.0022$ & \\ 
        $i=1$  & $0.2115\pm 0.0014$   &  $0.5262\pm 0.0287$  & $0.2027\pm 0.0622$   & $0.026\pm 0.039 $ & \\ 
        $i=2$  &  $1.802\pm 0.009$ & $-3.505\pm 0.154$    & $-2.310\pm 0.334$   &  $1.030\pm 0.210$ & \\ 
        $i=3$  & $-2.484\pm 0.018$  & $7.346 \pm 0.259$  & $3.033 \pm 0.562$   & $-3.083\pm 0.352$ & \\ \hline
        $\tilde{b}_{ij}$ & $ j=0 $ & $ j=1 $ & $ j=2$ & $ j=3 $& $j=4$ \\ \hline\hline
    & \multicolumn{5}{c}{Hovart anisotropy model }  \\ \hline
          $i=0$  & $(5.59\pm 0.11)\times 10^{-6}$ &$(1.38\pm 0.11)\times 10^{-5}$  & $(-1.24\times 0.26)\times 10^{-5}$ & $(-2.045\pm 0.51)\times 10^{-5}$& $(1.73\pm 0.34)\times 10^{-5}$ \\ 
          $i=1$  & $0.0018\pm 0.0002$ & $-0.0027\pm 0.0002$ &$0.0007\pm 0.00005$  & $0.0076\pm 0.001$ & $-0.0055\pm 0.0007$ \\ 
          $i=2$  & $0.0258\pm 0.0001$ &$0.0184\pm 0.001$  &$0.0028\pm 0.0074$ & $-0.063\pm 0.0074$  &$0.0434\pm 0.0051$ \\ 
          $i=3$  & $-0.058\pm 0.0003$ & $-0.0231\pm 0.0027$ & $-0.009\pm 0.006$ & $0.114\pm 0.012$ & $-0.077\pm 0.008$ \\  \hline
    & \multicolumn{5}{c}{Bowers Liang anisotropy model }  \\ \hline 
            $i=0$  &$(5.59\pm 0.11)\times 10^{-6}$   & $(-3.00\pm 0.19)\times 10^{-5}$  & $(-5.68\pm 0.47)\times 10^{-5}$ & $0.00011\pm 0.00009$  & $(-6.22\pm 0.61)\times 10^{-5}$ \\ 
          $i=1$  & $0.0018\pm 0.0002$ & $0.00231\pm 0.0003$ &$0.0108\pm 0.0009$ & $-0.0231\pm 0.0017$ &$0.0140\pm 0.0011$ \\ 
          $i=2$  & $0.0258\pm 0.0001$ & $-0.0175\pm 0.0031$ & $-0.0991\pm 0.0073$ & $0.1985\pm 0.0140$ & $-0.1202\pm 0.0096$\\ 
          $i=3$  & $-0.058\pm 0.0003$ & $0.0553\pm 0.0063$ &$0.1734\pm 0.0150$ & $-0.3915\pm 0.0285$ & $ 0.2412\pm 0.0195$ \\ \hline
\end{tabular}
\caption{Coefficients $\tilde{a}_{ij}$ and $\tilde{b}_{ij}$ for relations~\ref{eq:fit_MomegaR}, \ref{eq:fit_MomegaI} and \ref{eq:fit_coeffs}.}
\end{table*}

\section{Polar modes of anisotropic neutron stars}\label{sec:polarmodes}

Here, we restrict our analysis to the $\ell=2$ polar modes. We adopt the usual convention,  $\omega = 2\pi \nu + i/\tau$,  for the quasi-normal mode frequencies, with  with $\nu$ and $\tau$ denoting the oscillation frequency and damping time, respectively.

Figures~\ref{fig:omega} and \ref{fig:tau} show the oscillation frequency and damping time of the fundamental mode as functions of stellar mass for the three EOS considered and the two anisotropy prescriptions, namely the Horvat and Bowers-Liang models. The oscillation frequencies typically range from $1$ to $3$~kHz, while the damping times span approximately $0.5$--$1.25$~s, broadly consistent with other works \cite{Mondal:2023wwo, 2024PhRvD.110h3020L}, (we compare our formalism with other works  in Subsec. \ref{sec:comparison}). Regardless of the EOS or anisotropy prescription, the oscillation frequency generally increases with stellar mass, whereas the damping time decreases. For larger masses, increasing the anisotropy parameter, $\lambda$, toward more positive values generally leads to lower oscillation frequencies, with a more pronounced spread among the curves for the Bowers-Liang model. In the Horvat anisotropy model, higher anisotropy  leads to greater  oscillations frequencies at small masses.  The damping time is likewise sensitive to the anisotropy prescription: it decreases with increasing $\lambda$ in the Horvat model, whereas it increases with increasing $\lambda$ in the Bowers-Liang model.

\begin{figure}[t]
    \centering
    \includegraphics[width=0.99\columnwidth]{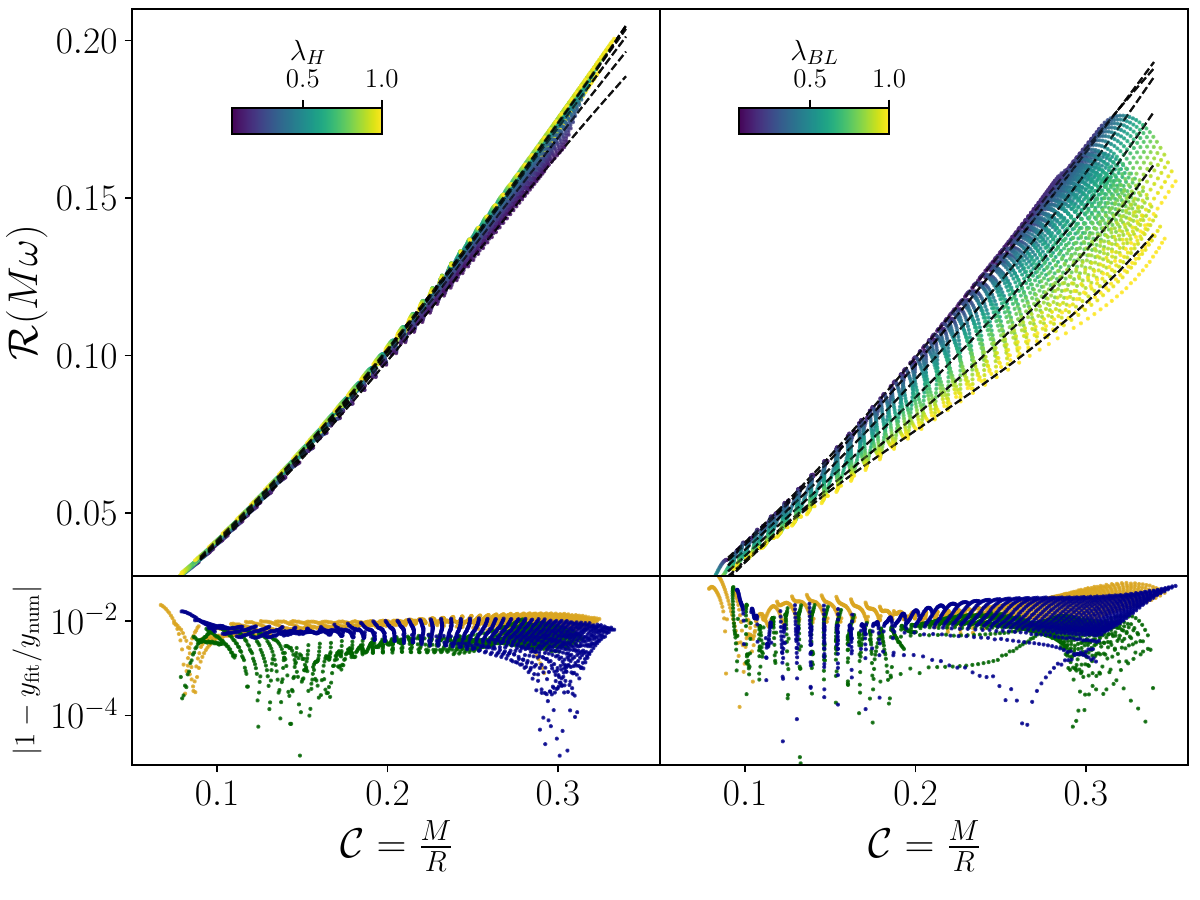}
    \includegraphics[width=0.99\columnwidth]{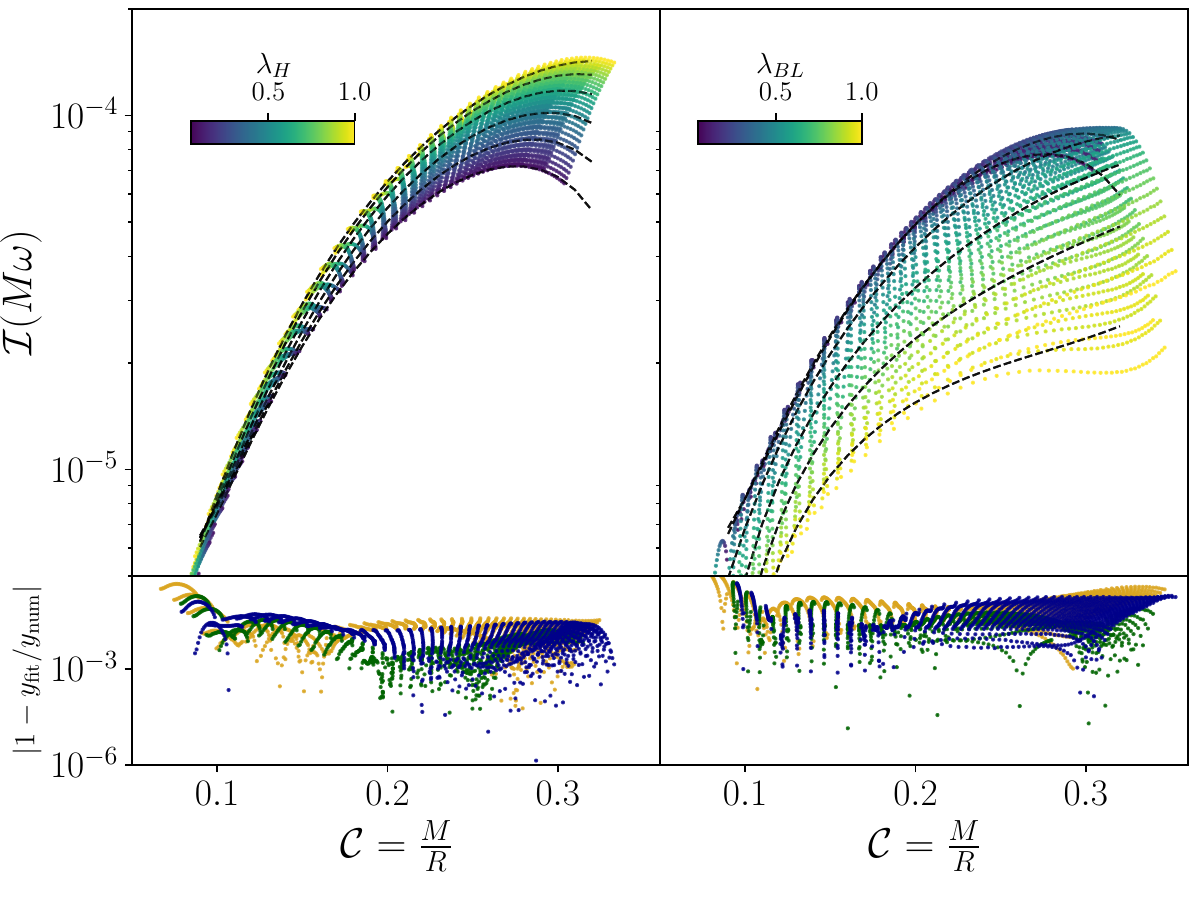}
    \caption{Dimensionless f-mode oscillation frequency $M\omega_R$ (upper panels) and damping time $M\omega_I$ (lower panels) as functions of the stellar compactness $\mathcal{C}=M/R$, for all three EOS and both anisotropy models: Horvat (left) and Bowers-Liang (right). The color scale indicates the value of the anisotropy parameter $\lambda$. The dashed black line shows the polynomial fit given by Eqs.~\eqref{eq:fit_MomegaR} and \eqref{eq:fit_MomegaI}. At the bottom of each panel, we show the relative error of the fit, defined as $|1 - y_{\rm fit}/y_{\rm num}|$, for each EOS: SLy4 (blue), GM1Y6 (green), and QHC21 (yellow). }
    \label{fig:modes_C}
\end{figure}

The upper and lower panels of Figure~\ref{fig:modes_C} show, respectively, the real and imaginary parts of the f-mode frequency, $M\omega_R$ and $M\omega_I$, as functions of the stellar compactness $\mathcal{C}=M/R$. In each panel, the dashed black line represents a polynomial fit of the form

\begin{equation}\label{eq:fit_MomegaR}
    M\omega_R(\mathcal{C}) = \sum_{i=0}^{3} a_i(\lambda)\,\mathcal{C}^i \,,
\end{equation}
\begin{equation}\label{eq:fit_MomegaI}
    M\omega_I(\mathcal{C}) = \sum_{i=0}^{3} b_i(\lambda)\,\mathcal{C}^i \,,
\end{equation}
where the coefficients $a_i$ and $b_i$ depend on the anisotropy parameter $\lambda$ and are themselves described by third-order polynomials,

\begin{equation}\label{eq:fit_coeffs}
    a_i(\lambda) = \sum_{j=0}^{3} \tilde{a}_{ij}\,\lambda^j \,, \qquad
    b_i(\lambda) = \sum_{j=0}^{4} \tilde{b}_{ij}\,\lambda^j \,.
\end{equation}

The coefficients $\tilde{a}_{ij}$ and $\tilde{b}_{ij}$ depend on the anisotropy model but are independent of the EOS, and are listed in Table~\ref{tab:fit_Momega}. In Figure~\ref{fig:modes_C} we also show the relative error of the fit proposed in each case, defined as $|1 - y_{\rm fit}/y_{\rm num}|$ for each EOS. In all cases, the relative error of the fit remains below $0.1$ .

\subsection{Comparison with other works}\label{sec:comparison}

\begin{figure}[h]
    \centering
    \includegraphics[width=0.79\columnwidth]{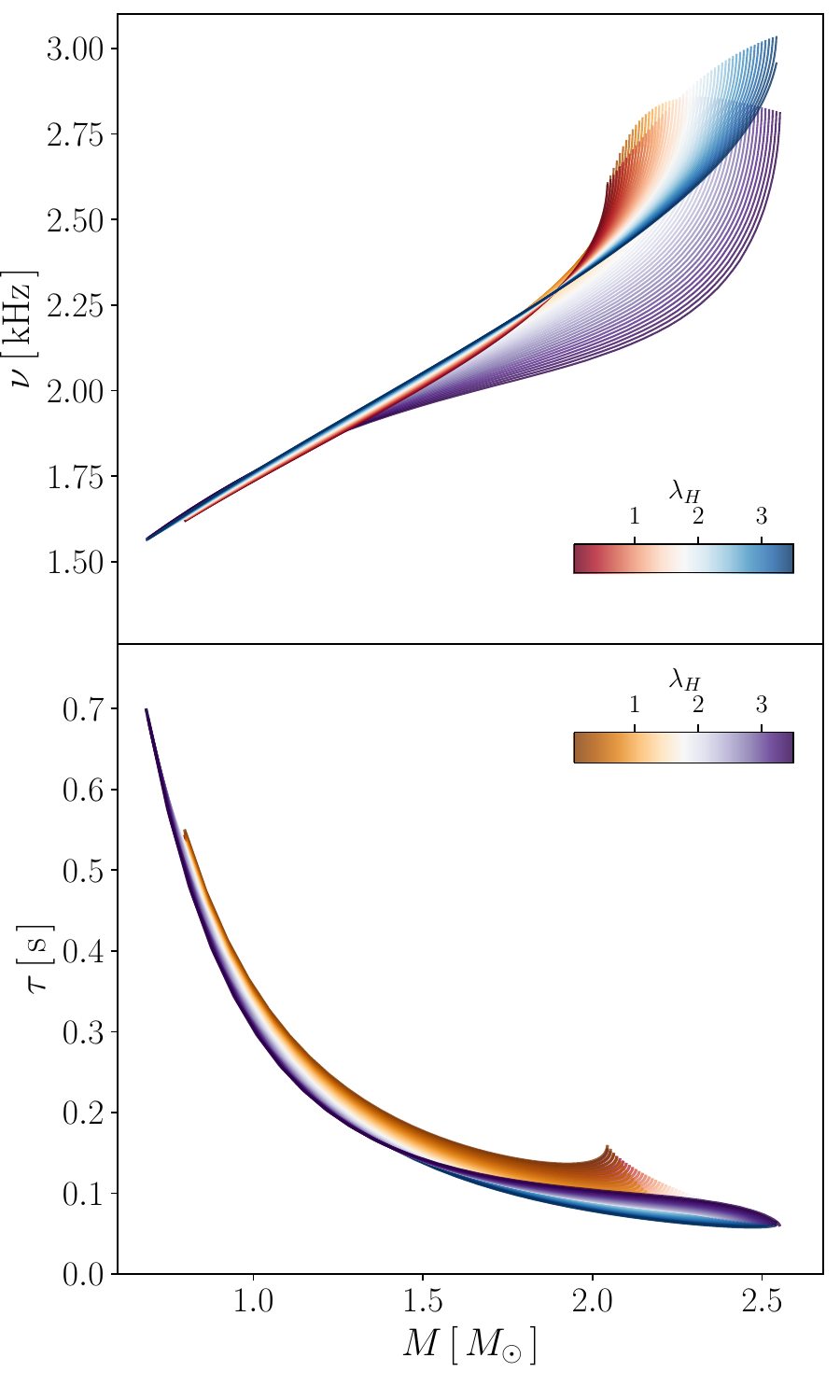}
    \caption{Oscillation frequency (top panel) and damping time (bottom panel) of the f-mode as a function of the stellar mass for the SLy4 EOS and the modified Horvat anisotropy model (given by equation~\ref{eq:sigma_H_modified}). The red-to-blue color bar corresponds to the case $\delta k^\theta=0$, while the orange-to-purple color bar corresponds to the case in which equation~\refeq{eq:Deltak} is imposed.}
    \label{fig:Comp}  
\end{figure}

\begin{table*}[]
    \centering
   \begin{tabular}{c|ccccccccc}   
     & $ \log_{10}(\rho_c)$ & $M/R$ & $\nu (\delta k^\theta= 0)$ &$\nu (\delta k^\theta\neq0)$&$\Delta \nu$ & $\tau (\delta k^\theta= 0)$& $\tau (\delta k^\theta\neq  0)$  & $\Delta \tau$ \\
     &  & & kHz & kHz & kHz & s & s & s \\ \hline
       & $14.782$ &  $0.099$  & $1.6252$ & $1.6247$ &  $-5.68\times 10^{-4} $ &  $0.5356$   &  $0.5303$  & $-5.33\times 10^{-3}$ 
      \\
      & $14.948$ & $0.177$ & $1.9730$ & $1.9662$ & $-6.84\times 10^{-3}$ & $0.1830$ & $0.1844$ & $1.43 \times 10^{-3}$ \\ 
     $\lambda=0.75 $ & $15.068$ & $0.233$ & $2.2378$ & $2.2209$ & $-0.0168$ & $0.1197$ & $0.1250$ & $5.31\times 10^{-3}$  \\ 
      & $15.162$ & $0.269$ & $2.4516$ & $2.4238$ & $-0.027$  & $0.1043$ & $0.1127$ &  $8.49\times 10^{-3}$ \\ 
      & $15.239$ & $0.291$ & $2.6368$ & $2.5938$ & $-0.043$ & $0.1040$ & $0.1164$ & $0.012$  \\ 
   \hline 
   & $14.772$  & $0.097$ & $1.6212$ & $1.6190$ & $-2.22\times 10^{-3}$ & $0.5474$  & $0.5441$ & $-3.30\times 10^{-3}$  \\ 
   & $14.939$ & $0.179$ & $1.9906$ & $1.9901$ & $-4.76\times 10^{-4}$ & $0.1607$ & $0.1684$ & $7.77\times 10^{-3}$  \\ 
   $\lambda=1.775$ & $15.059$   & $0.242$ & $2.2456$ & $2.2842$ & $0.0386$ &  $0.0985$ & $0.1031$ & $4.55\times 10^{-3}$  \\ 
   & $15.153$ & $0.285$ & $2.4603$ & $2.5243$ & $0.0639$ & $0.0828$ & $0.0873$ & $4.61\times 10^{-3}$  \\ 
   & $15.230$ & $0.311$ & $2.6703$ & $2.7262$ & $0.0559$ & $0.0777$ & $0.0884$ & $0.0106$ \\ 
   \hline
   & $14.753$ & $0.091$ &  $1.5951$ & $1.5913$ & $-3.87\times 10^{-3}$ & $0.6132$ & $0.6135$ & $-6.82\times 10^{-4}$ \\ 
   & $14.919 $ & $0.176$ & $1.9520$ & $1.9849$ & $0.032$ & $0.1594$ & $0.1652$ & $5.78 \times 10^{-3}$ \\ 
   $\lambda= 2.81$ & $15.039$ & $0.248$ & $2.1595$ & $2.3152$ &$0.1557$  & $0.0987$ & $0.0889$ & $-9.78\times 10^{-3}$ \\ 
   & $15.133$ & $0.299$ & $2.3579$ & $2.5890$ & $0.2311$ & $0.0829$ & $0.0697$ & $-0.01317$  \\ 
   & $15.211$ & $0.331$ & $2.6029$ & $2.8168$ & $0.2139$  & $0.0719$ & $0.0686$ & $-3.33\times 10^{-3}$ \\ 
   \hline

    \end{tabular}
    \caption{Comparison of the oscillation frequency  and damping time for the SLy4 EOS and the modified Horvat anisotropy model .}
    \label{tab:Omegadk0}
\end{table*}

\citet{2024PhRvD.110h3020L} also studied nonradial deformations of anisotropic neutron stars. The main difference between their approach and ours is that they do not impose equation~\eqref{eq:Deltak}; instead, the perturbation of $k^\alpha$ is constrained only by the orthogonality and normalization conditions. Specifically, our formalism reduces to theirs when the $\theta$-component of the Eulerian perturbation of $k^\alpha$ is neglected, $\delta k^\theta=0$. In this case, the system to integrate consists of four differential equations and two algebraic equations, with equation~\eqref{eq:polar_f} becoming an algebraic relation that directly determines the function $V$.

To ensure the regularity of the resulting equations, they further require $S_c=0$, which implies $\partial\bar{\sigma}x = \mathcal{O}(r^2)$. To satisfy this condition, they modify the Horvat anisotropy model as
\begin{equation}
\sigma_H = -\lambda_H(1-B)^2\bar{P}.
\label{eq:sigma_H_modified}
\end{equation}
In contrast, our formalism does not impose any such restrictions on the choice of anisotropy model.

In order to compare the two approaches on equal footing, we adopt, only in this section, the modified Horvat model in Eq.~\eqref{eq:sigma_H_modified}. This modification also changes the equilibrium configurations, widening the range for $\lambda_H$. The factor $(1-B)^2$ suppresses the anisotropy making the range for $\lambda_H$ wider. To guarantee causality for the radial and tangential sound speeds up to the maximum mass, we constrain the anisotropy parameter for the modified Horvat model to the range $0 \leq \lambda_H \leq 3.5$.  Figure~\ref{fig:Comp} compares the $f$-mode oscillation frequency and damping time as functions of the stellar mass for the SLy4 EOS, calculated using both formalisms.  The red-to-blue color bar corresponds to the case $\delta k^\theta=0$, which is consistent with the results of \cite{2026PhRvD.113d3025G}, while the orange-to-purple color bar corresponds to the case in which we impose equation~(\ref{eq:Deltak}). For small values of the anisotropy parameter, both formalisms yield very similar results, whereas their predictions begin to differ for larger values of the anisotropy. In particular, imposing equation~(\ref{eq:Deltak}) results in slightly lower $f$-mode frequencies and slightly longer damping times, especially for larger anisotropy parameters and higher stellar masses.

A more quantitative comparison is presented in Table~\ref{tab:Omegadk0}, where we report the oscillation frequencies and damping times obtained for the two cases, $\delta k^\theta=0$ and $\delta k^\theta\neq 0$, with the latter determined by equation~\eqref{eq:Deltak}. The comparison is performed for three values of the anisotropy parameter and several central mass densities (note that Fig. \ref{fig:Comp} compares at fixed stellar mass). While the differences in the oscillation frequencies remain below $10\%$, the damping times  can differ by up to $20\%$ for larger values of the anisotropy parameter.

\section{Discussions and Conclusions}\label{sec:discussion}
In this work, we have studied the effects of anisotropy on the $f$-modes of NS with realistic EOS. We used three representative EOS, QHC21, GM1Y6 and SLy4; and two anisotropy models, Bowers--Liang and Horvat. The Bowers-Liang and Horvat models describing $\bar{\sigma}$ do not specify how the anisotropy axis responds to a perturbation, so the perturbation problem is underdetermined and a prescription for the perturbation of $k^\alpha$ must be supplied.

In order to achieve this, we used relations involving the Lagrangian perturbations of the normalization and orthogonality between $u^\alpha$ and $k^\alpha$ to derive the perturbations of these vectors. Namely, we assumed that the \emph{Lagrangian} perturbation of $k^\alpha$ lies in the $(\bar{u}, \bar{k})$ plane, so that the corresponding Eulerian perturbation inherits an angular component. In contrast, other works have assumed that $\delta k^\theta = 0$, see e.g. \cite{PhysRevD.85.124023, Mondal:2023wwo, 2024PhRvD.110h3020L}, which is also consistent with the normalization and orthogonality relations and constrains instead the \emph{Eulerian} perturbation to the$(\bar{u}, \bar{k})$ plane, but implies that one degree of freedom of the fluid Lagrangian perturbation is not dynamical. The physical origin of the anisotropy, on the other hand, can be traced, for instance, to superfluidity or exotic phases of matter, i.e., it is tied to matter. Thus, we argue that if matter is disturbed, the anisotropy must be at least (Lie) dragged with it. This is the closure we have chosen here and in Ref.~\cite{2026PhRvD.114b4028R}.

The two prescriptions differ in the character of the tangential sector. When $\delta k^\theta = 0$ the $\theta$ component of $\delta(\nabla_{\mu}T^{\mu\nu}) = 0$ has no radial derivatives of $V$ and reduces to an algebraic constraint, whereas in our approach the term $\partial_r\left(\bar{\sigma}\partial_{r}V\right)$ appears in Eq.~\eqref{eq:TOVb}, and $V$ becomes dynamical. This equation resembles that of a shear wave with $\bar{\sigma}$ in the role of a \emph{shear modulus}. Since our prescription for the perturbation of the radial vector $k^\alpha$ is different, the Cowling limit does not coincide with one of the equations in Ref.~\cite{PhysRevD.85.124023}. The Cowling limit within our approach will be presented elsewhere.

Within this framework, we found that the $f$-mode oscillation frequency, which lies in the range $1$-$3$~kHz, increases with stellar mass, while the damping time, ranging from approximately $0.5$ to $1.25$~s, decreases with mass across the three EOSs and the two anisotropy prescriptions considered (Horvat and Bowers-Liang). Increasing the anisotropy, corresponding to tangential pressure exceeding radial pressure, generally lowers the oscillation frequency, while its effect on the damping time depends on the prescription: the damping time decreases with increasing anisotropy for the Horvat model but increases with increasing anisotropy for the Bowers-Liang model. We found that the real and imaginary parts of the scaled $f$-mode frequency, $M\omega_R$ and $M\omega_I$, satisfy quasi-universal relations with the stellar compactness $\mathcal{C}$ that are largely insensitive to the EOS. We provide simple polynomial fits to these relations, parametrized by the anisotropy parameter, with an accuracy better than $10\%$; these fits may serve as a useful tool for constraining neutron-star anisotropy through future asteroseismology observations.

Finally, the appearance of the new dynamical degree of freedom changes the behaviour of both the frequency and damping times of polar quasi-normal modes. This effect was also found for the axial $w$-modes of anisotropic neutron stars \cite{2026PhRvD.114b4028R}. Interestingly, the two prescriptions deviate appreciably when the anisotropy is large. This points to the effect of the shear modulus associated with the anisotropy in Eq.~\eqref{eq:TOVb}, i.e., the existence of the new degree of freedom, being significant, so that the two approaches can be distinguished by measurements of quasi-normal modes.  

\begin{acknowledgments}
 J.F.R-R is supported by the Universidad Antonio Nari\~no under grant
VCTI 2232026. F.D.L-C is supported by the Vicerrectoría de Investigación y Extensión - Universidad Industrial de Santander.
\end{acknowledgments} 

\appendix

\section{Equations }\label{app:I}

\subsection{Static Background Configuration}

The structure equations governing the unperturbed static background configuration are
\begin{eqnarray}
    \frac{rA'}{A} &=& -1+\frac{1}{B}+ \frac{8\pi r^2 \bar{P} }{B}\, , \label{eq:da} \\
      r B' &=& 1-B-8\pi r^2\bar{\epsilon}\, , \label{eq:db}
\end{eqnarray}
while the TOV equation takes the form
\begin{equation}\label{eq:dP}
    \bar{P}'=-(\bar{P}+\bar{\epsilon})\frac{A'}{2A}-\frac{2\bar{\sigma}}{r}\, .
\end{equation}
To ensure regularity at the stellar center, all background quantities are expanded in power series about $r=0$:
\begin{align}
    A &\approx A_c + A_2r^2+ A_4r^4+\mathcal{O}(r^6),\\
B & \approx1 - \frac{8\pi\bar{\epsilon}_c}{3}r^2-\frac{8\pi\bar{\epsilon}_2}{5}r^4+\mathcal{O}(r^6),\\
\bar{P} & \approx  \bar{P}_c + \bar{P}_2r^2+\bar{P}_4r^4+\mathcal{O}(r^6),
\end{align}
where the subscript $c$ denotes quantities evaluated at the stellar center. The expansion coefficients satisfy
\begin{align}
    A_2 &= 4\pi\left(\bar{P}_c+\frac{\bar{\epsilon}_c}{3}\right),\\
    A_4 &= 2\pi A_c  \left[ \bar{P}_2 + \frac{\bar{\epsilon}_2}{5}  + 
   4 \pi\left(\bar{P}_c + \bar{\epsilon}_c\right) \left( \bar{P}_c + \frac{\bar{\epsilon}_c}{3}\right) \right],\\
   P_2 &=  -\frac{A_2}{2A_c}(\bar{P_c}+\bar{\epsilon}_c)-\bar{\sigma}_2,\\
   P_4 &= \frac{(\bar{P}_c+\bar{\epsilon}_c)}{2}\left(\frac{A_2^2}{2A_c^2} -\frac{A_4}{A_c}\right) -\frac{\bar{\sigma}_4}{2}-\frac{A_2}{4A_c}(\bar{P}_2+\bar{\epsilon}_2),\\
   \bar{\epsilon}_2 &= \bar{P}_2 \left(\frac{d\bar{\epsilon}}{d\bar{P}}\right)_c.
\end{align}

The anisotropy is likewise expanded around the stellar center as
\begin{equation}
    \bar{\sigma}\approx \bar{\sigma}_2 r^2 + \bar{\sigma}_4 r^4 + \bar{\sigma}_6 r^6+\mathcal{O}(r^8).
\end{equation}
For the Bowers--Liang anisotropy model, the leading coefficients are
\begin{align}
    \sigma_2^{BL}&=-\frac{\lambda_{BL}}{3}(\bar{P}_c+\bar{\epsilon}_c)\left(\bar{P}_c+\frac{\bar{\epsilon}_c}{3}\right), \\
    \sigma_4^{BL}&=-2\lambda_{BL}\left[\bar{\epsilon}_2\left(2\bar{P}_c+\bar{\epsilon}_c\right) +\bar{P}_2(3\bar{P}_c+2\bar{\epsilon}_c)+\right.\nonumber \\
    &\left.\frac{4\pi} {3}\bar{\epsilon}_c(\bar{P}_c+\bar{\epsilon}_c)(3\bar{P}_c+\bar{\epsilon}_c)\right].
\end{align}
For the Horvat anisotropy model, the corresponding coefficients are
\begin{align}
    \sigma_{2}^H&=-\lambda_H \left(\frac{ 8\pi\bar{\epsilon}_c}{3}\right)\bar{P}_c,\\
    \sigma_4^{H}&=-\lambda_H \left(\frac{ 8\pi\bar{\epsilon}_c}{3}\right)\bar{P}_c\left(\frac{1}{5}\frac{\bar{\epsilon}_2}{\bar{\epsilon}_c}+\frac{\bar{P}_2}{\bar{P}_c}\right).
\end{align}

\subsection{Polar perturbation equations}

To improve the numerical behavior, and assuming a harmonic time dependence for the perturbation variables, we perform the following change of variables: 
\begin{align}
    H_0 &\rightarrow r^{\ell } H_0(r) e^{i \omega t},\\ H_1 &\rightarrow i \omega r^{\ell +1} H_1(r) e^{i \omega t}, \\
    K &\rightarrow r^{\ell} K(r) e^{i \omega t},\\
    W &\rightarrow  r^{\ell-1} W(r) e^{i \omega t},\\
    V &\rightarrow r^{\ell-2} V(r) e^{i \omega t},\\
    \delta P &\rightarrow P_2(r) e^{i \omega t}, \\
    \delta \epsilon &\rightarrow \epsilon_2(r) e^{i \omega t}. 
\end{align}
In terms of the new variables, the system governing the polar perturbations of anisotropic configurations, Eqs.~\ref{eq:EEa}, \ref{eq:EEb}, \ref{eq:TOVa}, \ref{eq:TOVb}, and \ref{eq:TOVc}, takes the form
\begin{multline}\label{eq:polar_a}
    rH_1' = \frac{H_0+K}{B} - \frac{H_1}{B}\left( 4\pi r^2(\bar{P}-\bar{\epsilon})+\ell B +1\right)\\+\frac{16\pi V}{ B}\left(\bar{P}+\bar{\epsilon}-\bar{\sigma}\right),
\end{multline}
\begin{multline}\label{eq:polar_b}
    rK'= H_0+\frac{\ell(\ell+1)H_1}{2}-K\left( 1+\ell - \frac{rA'}{2A} \right)\\ +\frac{8\pi W}{B}\left(\bar{P}+\bar{\epsilon}\right),
\end{multline}

\begin{multline}\label{eq:polar_c}
    rW'=-r^2\left(\frac{H_0}{2}- \frac{\tilde{X}}{\sqrt{A}\Gamma_1\bar{P}}\right) 
    +W\left( 1-\ell + \frac{rB'}{2B} \right) \\  -(1 - \tilde{\sigma}) \left(r^2K +2W+\ell(\ell+1) V\right),
\end{multline}

\begin{equation}\label{eq:polar_d}
       rV' = -\frac{r\chi}{\bar{\sigma}} -(\ell-2) V,
\end{equation}

\begin{multline}\label{eq:polar_e}
   r\tilde{X}'=\sqrt{A}(\bar{\epsilon}+\bar{P})\left[\left( -\ell(\ell+1)\left(\frac{1}{2}-\tilde{\sigma}\right) -\frac{r^2\omega^2}{A}\right)\frac{H_1}{2}\right.\\+\left(4\tilde{\sigma}^2-6\tilde{\sigma}  +(2\tilde{\sigma}-1)\frac{rA'}{A}\right)\frac{K}{2} \\-\left(1-\frac{rA'}{2A}\right)\left(\frac{K}{2}-\frac{H_0}{2}\right)  - 8\pi r \chi \\
    + \left( \frac{  r^2}{2\sqrt{B} }\left(\frac{\sqrt{B}A'}{r^2A}\right)' -\frac{\omega^2}{AB}  -\frac{4\pi (\epsilon+P)}{B} \right)W\\
  \left.  \left(\frac{2\tilde{\sigma}}{r}  + \frac{A'}{A}-\frac{3}{r}+\frac{r\bar{\sigma}'}{\bar{\sigma} }\right) \frac{2\tilde{\sigma}}{r}W \right] +\left(\frac{2\bar{\sigma}}{\Gamma_1\bar{P}}-\ell\right)\tilde{X} \\+ \sqrt{A}\left( 2S +\frac{\ell(\ell+1)}{r}\left( \bar{P}'\left(1-\tilde{\sigma}\right)V -\frac{\chi}{r}\right) \right),
\end{multline}

\begin{multline}\label{eq:polar_f}
    \chi'= \frac{rS}{B} +\frac{r\tilde{X}}{\sqrt{A}B}+\frac{r\left(\bar{\epsilon}+\bar{P}\right)}{B}\left[\frac{H_0}{2}-\frac{ (1-\tilde{\sigma})  \omega^2 V}{A}\right] \\+\frac{W p'}{B} - \left( \frac{1 + \ell}{r} + \frac{A'}{2A}+ \frac{B'}{2B} \right)\chi ,
\end{multline} 

as well as the algebraic constraint given by Eq.~\ref{eq:EEe}:
\begin{multline}\label{eq:polar_g}
    \frac{8\pi r^2 \tilde{X}}{\sqrt{A}} +\left(\ell(\ell+1)-2B +\frac{rBA'}{A}\right)\frac{H_0}{2} \\ - 16\pi\bar{\sigma}W - B\chi   + B\left(\frac{\ell(\ell+1)rA'}{4A}-\frac{r^2\omega^2}{A}\right)H_1 \\+\left[1-\frac{1}{2}\ell(\ell+1)+\frac{r^2\omega^2}{A}+\frac{rBA'}{2A}\left(1-\frac{rA'}{2A} \right)\right]K=0.
\end{multline}

We futher define :
\begin{align}
    \tilde{\sigma}&=\frac{\bar{\sigma}}{\bar{\epsilon}+\bar{P}},\\
    \Delta P &=  \frac{r^\ell \tilde{X} e^{i \omega t} }{\sqrt{A}} Y^{\ell 0}, \\
    \delta \sigma &= r^\ell S e^{i \omega t}  Y^{\ell 0} , 
\end{align}
with $\Delta P$ denotes the Lagrangian perturbation of the pressure, $\delta\sigma$ is the Eulerian perturbation of the anisotropy, and $\tilde{\sigma}$ is the dimensionless anisotropy parameter.

The boundary conditions at the stellar center are obtained by expanding each perturbation variable in a power series about $r=0$. For any perturbation variable $\Psi=\{H_0, H_1, K, W, V, \tilde{X}\}$, we write
\begin{equation}
\Psi(r)=\Psi_c+\Psi_2 r^2+\mathcal{O}(r^4),
\end{equation}
and 
\begin{equation}
\chi(r)=\chi_2r^2+\chi_4r^4+\mathcal{O}(r^6),
\end{equation}
Substituting these series into the perturbation equations and matching equal powers of r yields the expansion coefficients recursively. At leading order, we obtain
\begin{align}
    H_{0,c} &= K_{c},\\
    H_{1,c} & = \frac{2}{1+\ell}K_c +\frac{16\pi(\bar{P}_c+\bar{\epsilon}_c)}{1+\ell}V_c,\\
    W_c &= -\ell V_c, \\
    \tilde{X}_c & = \sqrt{A_c}\left[(\bar{P}_c+\bar{\epsilon}_c)\left(\frac{\omega^2 }{A_c}V_c-(1+\ell)\chi_2-\frac{K_c}{2} \right) \right. \nonumber \\
        & \left.+  \frac{2S_c}{\ell} +2\ell \bar{P}_2V_c  \right], \\
    S_c  &= \left(1+\frac{2}{\ell}\left(\frac{\partial \bar{\sigma}}{\partial \bar{P}}\right)_c\right)^{-1}\left[ K_c\left(\frac{\partial \bar{\sigma}}{\partial \mu}\right)_c \right. \nonumber\\
    & \left. + (\bar{P}_c+\bar{\epsilon}_c)\left(\frac{K_c}{2}+(1+\ell)\chi_2-\frac{\omega^2}{A_c} V_c\right)\left(\frac{\partial \bar{\sigma}}{\partial \bar{P}}\right)_c \, \right].
\end{align}
The next-order coefficients satisfy
\begin{align}
    \chi_2 &= -(\ell-2)\bar{\sigma}_2 V_c \\
    V_2 &= (2+\ell)\frac{S_c}{2\ell}-(\ell+1)(1+\bar{P}_c+\bar{\epsilon}_c)\frac{\chi_2}{2}\\
    \chi_4 &= (2-\ell)\bar{\sigma}_4V_c-\ell\bar{\sigma}_2V_2, 
\end{align}
with $\mu=1-B$. The remaining second-order coefficients, $\{H_{0,2}, H_{1,2},W_2,X_2.S_2\}$, are determined by solving the following system of algebraic equations:
\begin{multline}
     S_2 + \left(2\bar{P}_2W_2 +\frac{\tilde{X}_2}{\sqrt{A_c}}\right)\left(\frac{\partial \bar{\sigma}}{\partial \bar{P}}\right)_c\\ - H_{0,2} \left(\frac{\partial \bar{\sigma}}{\partial \mu}\right)_c = \frac{\tilde{X}_c}{\sqrt{A_c}}\left[ \frac{A_2}{2A_c}\left(\frac{\partial \bar{\sigma}}{\partial \bar{P}}\right)_c-\left(\frac{\partial^3 \bar{\sigma}}{\partial^3 P}\right)_c\right]   \\
      -H_{0,c}\left[B_2\left(\frac{\partial \bar{\sigma}}{\partial \mu}\right)_c + \left(\frac{\partial^3 \bar{\sigma}}{\partial^3 \mu}\right)_c\right] \\ -2W_c \left[2\bar{P}_4\left(\frac{\partial \bar{\sigma}}{\partial \bar{P}}\right)_c + \bar{P}_2\left(\frac{\partial^3 \bar{\sigma}}{\partial^3 \bar{P}}\right)_c\right],
\end{multline}

\begin{multline}
     H_{0,2}+K_{2}-(1+\ell)H_{1,2} = B_2(H_{1,c}-2K_c)-4\pi(\bar{P}_c-\bar{\epsilon}_c)H_{1,c}\\-2V_2+16\pi(\bar{P}_c+\bar{\epsilon}_c)( V_2 - B_2V_c + (\bar{P}_2+\bar{\epsilon}_2-\bar{\sigma}_2)V_c),
\end{multline}

\begin{multline}
    (\ell^2+\ell-2)(H_{0,2}-K_2)= 32\pi\left(\sigma_2W_c+\chi_2-\frac{X_c}{2\sqrt{A_c}}\right)\\-\left(\ell(\ell+1)\frac{A_2}{A_c}-\frac{2\omega^2}{A_c}\right)H_{1,c}+2\left(B_2-\frac{\omega^2}{A_c}\right)K_{1,c},
\end{multline}

\begin{multline}
    (1+\ell)\left(K_2-\frac{\ell}{2}H_{1,2}\right)-H_{0,2}-8\pi(P_c+\epsilon_c)W_2= -2V_2\\+8\pi\left[\left(\bar{P}_c+\frac{\bar{\epsilon}_c}{3}\right)\frac{K_c}{2}+\left(\bar{P}_2+\bar{\epsilon}_2-B_2(\bar{P}_c+\bar{\epsilon}_c)\right)W_c\right],
\end{multline}

\begin{multline}
  (1+\ell)W_2 =  -\frac{H_{0,c}}{2}-K_c-(2+\ell+\ell^2)V_2 - B_2W_c \\\ +\frac{P_2\tilde{X}_c}{\sqrt{A_c}\bar{\epsilon}_c}  + \ell(\ell-1)\tilde{\sigma}_2V_c,
\end{multline}

\begin{multline}
    \frac{H_{0,2}}{2}-\frac{K_2}{2}+\frac{\ell(\ell+1)}{4}H_{1,2}+\frac{\ell \tilde{X}_2}{\sqrt{A_c}(\bar{P}_c+\bar{\epsilon}_c)}\\ + \left(8\pi\left(\bar{P}_c+\frac{2\bar{\epsilon}_c}{3}\right) +2\tilde{\sigma}_2+\omega^2 \right)W_2-\frac{2S_2}{\bar{P}_c+\bar{\epsilon}_c} =\\
    -\frac{A_2}{A_c}\left( \ell(\ell+1)\left(\frac{\chi_2}{2}+V_2\right)+(4+\ell)\frac{K_c}{4} - \frac{S_c}{\bar{P}_c+\bar{\epsilon}_c}+\frac{\ell\omega^2}{2A_c}V_c\right)\\
     -  \frac{\ell(\bar{P}_2 + \bar{\epsilon}_2) }{\bar{P}_c + \bar{\epsilon}_c } \left( (\ell+1) \chi_2 +\frac{K_c}{2} +\left(\ell\frac{A_2}{A_c}+\frac{\omega^2}{A_c}\right)V_c\right)\\
    -\ell  (  \ell+1 )\chi_4 - 8\pi ( \bar{P}_c + \bar{\epsilon}_c )\left( \chi_2+\frac{\omega^2V_c}{A_c(\ell+1)}\right) + \frac{2 \bar{\sigma}_2\bar{P}_2}{\bar{\epsilon}_2}\frac{\tilde{X}_c}{\sqrt{A_c}} \\
  - \ell\tilde{\sigma}_2\left( 2(\ell+1)V_2 +(3-\ell)\frac{K_c}{\ell}+4\pi\left(\bar{P}_c-\bar{\epsilon}_c\right) V_c\right)  \\
        - \frac{2V_2}{\sqrt{A_c}(\bar{P}_c+\bar{\epsilon}_c)} 
                  -\frac{\omega^2}{A_c(1+\ell)} \left(  K_c +  (2-\ell-\ell^2)B_2V_c   \right)\\
  + \ell V_c\left(  2(\ell+2)\frac{A_4}{A_c} -\frac{(\ell+4)}{2}\left(\frac{A_2}{A_c}\right)^2 + \frac{A_2B_2}{A_c} \right)  - \\
  2\ell V_c \left(  2\pi(\bar{P}_2+\bar{\epsilon}_2-B_2(\bar{P}_c+\bar{\epsilon}_c))-(\ell-1)\tilde{\sigma}_2^2+(\ell+2)\tilde{\sigma}_4\right).
\end{multline}


\bibliography{apssamp}

\end{document}